\documentstyle[12pt]{article}\textheight 230mm\textwidth 160mm
\newcommand{\bi}{\bibitem}
\newcommand{\be}{\begin{eqnarray}}
\newcommand{\ee}{\end{eqnarray}}
\newcommand{\nn}{\nonumber}
\newcommand{\al}{\alpha}
\newcommand{\bt}{\beta}

\newcommand{\de}{\delta}
\newcommand{\la}{\lambda}

\newcommand{\ol}{\overline}
\begin{document}
\hspace*{11.6cm}KANAZAWA-93-06
\vspace*{1cm}

\begin{center}
 {\bf Weyl Invariance and Spurious Black Hole \\
in \\
Two-Dimensional Dilaton Gravity}
\end{center}

\vspace*{1cm}
\begin{center}{Takanori Fujiwara$\ ^{(1),(a)}$, Yuji Igarashi$\ ^{(2),(b)}$
and  Jisuke Kubo$\ ^{(3),(c)}$}
\end{center}
\vspace*{0.2cm}
\begin{center}
{\em $\ ^{(1)}$ Department of Physics, Ibaraki University, Mito 310, Japan}\\
{\em $\ ^{(2)}$Faculty of General Education, Niigata
University, Niigata 950-21, Japan}\\
{\em $\ ^{(3)}$ College of Liberal Arts, Kanazawa University, Kanazawa 920,
Japan }
\end{center}
\vspace*{0.5cm}
\begin{center}
ABSTRACT
\end{center}
In two-dimensional dilaton gravity  theories, there
may exist a global Weyl invariance which makes black hole
spurious. If the global invariance and
the local Weyl invariance
of the matter coupling are
intact at the quantum level, there is no Hawking radiation.
We explicitly verify the absence of anomalies in these symmetries
for the model proposed by
Callan, Giddings, Harvey
and Strominger.
The crucial observation is that
the conformal anomaly can be cohomologically trivial
and so not truly anomalous
in such dilaton gravity models.
\vspace*{1cm}
\footnoterule
\vspace*{4mm}
\noindent
$^{(a)}$ E-mail address: tfjiwa@tansei.cc.u-tokyo.ac.jp  \\
$^{(b)}$ E-mail address: igarashi@ed.niigata-u.ac.jp\\
$^{(c)}$ E-mail address: jik@hep.s.kanazawa-u.ac.jp
\newpage
\pagestyle{plain}
\section{Introduction}
The two-dimensional cosmological model of Callan, Giddings, Harvey
and Strominger (CGHS)
\cite{cghs} \footnote{See ref. \cite{HS}
and references therein for review
and recent developments.}
 has two candidates, $g_{\al\bt}$ and
$\hat{g}_{\al\bt}$, for the metric.
\footnote{$\hat{g}_{\al\bt}$ is defined in eq. (10) in the next section.}
In the world
described by $g_{\al\bt}$, there exists a black hole,
but it disappears in the $\hat{g}_{\al\bt}$-world \cite{cghs}.
It seems at first sight that one can choose one
of them for one's own purpose.
However, there is a global symmetry \cite{su}
with the transformation acting on $g_{\al\bt}$ and $\hat{g}_{\al\bt}$
differently, and one
finds
that the curvature scalar $R(g)$
experiences a change under the global
symmetry transformation while $R(\hat{g})$ does not.
So one may be naturally led to choose $\hat{g}_{\al\bt}$
as the metric
because $R(\hat{g})$ has the absolute
meaning with respect to the symmetry transformation.
As we will see in
section 2, the global symmetry is a Weyl symmetry, and is
responsible also for
decoupling of the matter fields from the conformal factor of $g_{\al\bt}$,
implying that the matter does not feel the existence of the black hole.
This suggests that physics remains the same whether one chooses
$g_{\al\bt}$ or $\hat{g}_{\al\bt}$ as the metric.

Quantum theoretically things may be different because Weyl symmetry
is anomalous in general \cite{deser,PolA}. In ref. \cite{fik} (hereafter
referred to as I), we have investigated the algebraic structure of
possible anomalies in the CGHS model, by considering the consistency
conditions on anomalies. Under a certain set of reasonable
assumptions, we have shown that there exist no cohomologically
non-trivial anomalies. This result of the pure algebraic
investigation has motivated us to confirm
the absence of cohomologically non-trivial anomalies by
explicit calculation of anomalies.

 There may be many physically inequivalent
 quantum theories
to a single classical theory. We will be assuming
in this paper that they have
at least following two properties:
\newline
(i): In the limit $\hbar \to 0$, they should approach to
the classical theory in some definite sense.
\newline
(ii): All the symmetries of the classical theory have to be realized
in the quantum theories in a corresponding fashion, except
for those which can not be realized because of anomaly,
 but does not influence
the consistency of the theories (scale invariance
in QCD for instance).
What we would like to attempt to explicitly show
in subsequent sections is
 that,
to the classical
CGHS model,
there are no quantum theories in which the CGHS black holes
and the corresponding
Hawking radiations are physically real, rather than that
there is one quantum theory in which the CGHS black holes are
spurious. Obviously, this finding is in a sharp contrast to the results
of
the previous investigations on the CGHS black holes.

In section 3 we will briefly review the canonical setting
which is based on the extended-phase-space method of Batalin, Fradkin and
Vilkovisky (BFV) \cite{bfv}. The conformal-gauge equivalent
of the algebraic proof on the absence of anomalies of I
will be presented in section 4.

We will start, in section 5, to calculate the contributions to
anomalous Schwinger terms. We would like to apply the results of Polyakov
\cite{PolA} and Fujikawa \cite{fujikawa} on the conformal anomaly to
determine the contribution of the matter fields. To use their results
appropriately, we will be assuming in that section that
the metric which
couples to the matter ($\ol{g}_{\al\bt}$) differs
from $g_{\al\bt}$
in the conformal factor. Classically, there is no difference,
because the conformal factor of $\ol{g}_{\al\bt}$ does not appear
in the action thanks to the local Weyl invariance of
the matter coupling.
We will show that this local Weyl symmetry is intact
at the quantum level. This implies, on one hand, that
the theory with $\ol{g}_{\al\bt}$ in the
matter coupling is gauge-equivalent to the original theory,
and on the other hand, that the conformal factor of $g_{al\bt}$
really decouples from the matter.

Furthermore, the matter
does not contribute to the anomaly in the global Weyl invariance
in this modified system
because it is the invariance under the change of
$g_{\al\bt}$ and not of $\ol{g}_{\al\bt}$. So, the matter
sector has nothing
to do with the anomalies in the CGHS model.
This is why we will consider the model without the matter fields
in the last part of section 5.
We will go to the modified light-cone gauge, the light-cone
gauge for $\hat{g}_{\al\bt}$ \cite{terao}, in which the theory
takes the most simple form, and find no anomalies,
conforming the previous, algebraic result. We will also discuss
how the result
can be modified if the system
in the modified light-cone gauge couples to
conformal matter. We will find again that the matter contribution
to anomalies is cohomologically trivial and so not really anomalous.

Section 6 is devoted to discussion and appendix
contains an algebraic proof which is needed in section 4.
\section{Spurious black hole in the classical approximation}
The CGHS action
is given by \cite{cghs}
\be
S^{\rm cl} &=& \int d^{2}\sigma\ \sqrt{-g}\,\{\,\exp (-2\phi)\,[\,
R(g)+4\,g^{\alpha\beta}\partial_{\alpha}\phi
\,\partial_{\beta}\phi +4\,\mu^2\,]\nn\\
& &-\frac{1}{2}\sum_{i=1}^{n} g^{\alpha\beta}
\ \partial_{\alpha}f_{i}
\,\partial_{\beta}f_{i}\,\}\ ,\ (~\alpha,\beta~=~0,1~)\ ,
\ee
where $f's$ stand for matter fields, $\phi$ for the dilaton field,
$\mu^2$ for a cosmological constant, and the curvature scalar
for the metric $g_{\alpha\beta}$ is denoted by $R(g)$.
The action is invariant under
2D general coordinate transformations, and
under the global, non-linear transformation \cite{su,fik}
\be
\phi &\rightarrow& \phi '=
\phi-\frac{1}{2}\ln (1+\Lambda\exp(2\phi)\,)~,\nn\\
g_{\alpha\beta} &\rightarrow&g_{\alpha\beta} '\,=\,
g_{\alpha\beta}(1+\Lambda\exp (2\phi) )^{-1}~,
\ee
where $\Lambda$ is a constant parameter.
This global symmetry manifests itself in degeneracy of classical
solutions, and is responsible for the unrecognizability
of the black hole in the model as we have stated in I.

To see this, let us  work in the conformal gauge,
\be
-g_{00}&=&g_{11}~=~=\exp 2\rho~,~ g_{01}~=~0~,
\ee
and write the action (1) in this gauge:
\footnote{The Lagrangian for the action (4) differs from that
of (1) by a total derivative. But the canonical structure
derived from (4) agrees with that of the next section.}
\be
S^{\rm cl}_{\rm CG} &=&
\int d^2\sigma\,\{~-\partial_{+}\psi\,\partial_{-}(
\rho-\phi)
-\partial_{-}\psi\,\partial_{+}(
\rho-\phi)\nn \\
& &+4\,\mu^2\,\exp 2(\rho-\phi)
+\frac{1}{2}\,\sum_{i=1}^{n}\,\partial_{+}f_{i}\partial_{-}f_{i}~\}
{}~,\\
\mbox{with} & & \psi \equiv \exp(-2\phi)~,\nonumber
\ee
where we have used $\partial_{\pm}=\partial_{0}\pm\partial_{1}$ so that
$\partial_{\pm}\,\sigma^{\pm}=2$, and we will also use
the abbreviations, $ \dot{h} = \partial_{0} h
=\partial_{\tau} h~, ~
h^{\prime} = {\partial}_{1}h ={\partial}_{\sigma}h$.
The action yields the equations of motion
\be
\partial_{+}\partial_{-}(\rho-\phi) &=&0~,~
\partial_{+}\partial_{-}\psi+4\mu^2 \,\exp2(\rho-\phi) ~=~0~.
\ee

Under
the transformation defined in (2), $\rho$ transforms as
\be
\rho '&\rightarrow&\rho-\frac{1}{2}\ln (1+\Lambda\exp(2\phi)\,)~,
\ee
so that the black
hole solution \cite{cghs},
\be
\rho &=&\phi=\rho_{\rm BH}~,~
\exp (-2\rho_{\rm BH}) ~=~ \frac{M}{\mu}- \mu^2 \sigma^+\sigma^-~,
\ee
undergoes the change
\be
\exp -2\rho_{\rm BH} ^{\prime} &=&\exp -2\rho_{\rm BH}+\Lambda~.
\ee
By applying the symmetry transformation appropriately, we can arrive at
\be
\exp -2\rho_{\rm BH} ^{\prime\prime} &=&-\mu^2\,
\sigma^{+}\sigma^{-}~,~~
\mbox{with}\, \,R(\rho_{\rm BH} ^{\prime\prime}) ~=~0~.\nn
\ee
This sounds strange because the curvature
has no longer the absolute meaning.
In fact, the curvature scalar $R(g)$ non-trivially
transforms
under the transformation (2):
\be
\sqrt{-g'}\,R(g') &=&\sqrt{-g}\,R(g)+\partial_{\al}[\,
\sqrt{-g}g^{\al\bt}\partial_{\bt}\,\ln (1+\Lambda\exp2\phi)\,]~.
\ee
In this connection, we would like to mention that the re-scaled quantity
\be
\hat{g}_{\alpha\beta}\equiv g_{\alpha\beta}\exp (-2\phi)
\ee
can equally well be used as the two-dimensional metric.
So, at the classical level, there are at least two candidates
for the metric. But the discussion above suggests that we should
presumably
regard $\hat{g}_{\alpha\beta}$ as the true metric because
it does not change under the global transformation (2)
and hence has the absolute meaning. Needless to say that
in that case the classical back ground (7) does not exhibit
a black hole any more.
In the following discussions, we nevertheless consider
$g_{\al\bt}$ as the metric and quantize the theory.
Our result however suggests that physics remains the
same whether we regard $g_{\alpha\beta}$  or $\hat{g}_{\alpha\beta}$ as the
metric.
 \section{The Batalin-Fradkin-Vilkovisky quantization}
\subsection{Parametrization and constraints}
To quantize the theory,
we employ
the generalized Hamiltonian method
of BFV \cite{bfv}, and
use the Arnowitt-Deser-Misner parametrization
of the metric $g_{\alpha\beta}$:
\be
g_{\alpha\beta}&\equiv&
\left( \begin{array}{cc} -\lambda^{+}\lambda^{-} & (\lambda^{+}
-\lambda^{-})/2 \\
(\lambda^{+}-\lambda^{-})/2 & 1 \end{array} \right)\exp 2\rho ~.
\ee
In terms of these new variables, the CGHS action becomes
\be
S^{\rm cl} &=&\int d^2 \sigma\,(~
{\cal L}^{\rm cl}_{{\rm d}}+{\cal L}^{\rm cl}_{\mu}+
{\cal L}^{\rm cl}_{f}~) ~,
\ee
where
\be
{\cal L}^{\rm cl}_{{\rm d}}&=&\frac{\psi '}{\lambda^+
+\lambda^-}[\, 2\,(\lambda^+ -\lambda^-)(\dot{\rho}-\dot{\phi})
+4\,\lambda^+\lambda^-(\rho '-\phi ')
+2\,(\lambda^+ \lambda^-) '\,]\nn\\
& &+\frac{\dot{\psi}}{\lambda^+ +\lambda^-}[\,
-4\,(\dot{\rho}-\dot{\phi})
+2\,(\lambda^+ -\lambda^-)(\rho '-\phi ')
+2\,(\lambda^+ -\lambda^-) '\,]~,\nn\\
{\cal L}^{\rm cl}_{\mu}&=&2\mu^2\,
(\lambda^+ +\lambda^-)\exp 2(\rho-\phi)~,\nn\\
{\cal L}^{\rm cl}_{f}&=&
\frac{1}{\lambda^+ +\lambda^-}\sum_{i=1}^{n}\,[\,
\dot{f}_{i}\dot{f}_{i}
-(\lambda^+ -\lambda^-)\dot{f}_{i}f_{i} '
-(\lambda^+ \lambda^-)f_{i} 'f_{i} ' \,]~.\nn
\ee
The conjugate momenta to $\lambda^{\pm},\rho,\phi$ and $f_{i}$
are  respectively given by
\be
\pi_{+}^{\lambda} &=&0~,~\pi_{-}^{\lambda}=0~,\nn\\
\pi_{\rho} &=& (\lambda^+ +\lambda^-)^{-1}\,[\,
2\,(\lambda^+ -\lambda^-)\psi '-4\dot{\psi}\,]~,\nn\\
\pi_{\phi} &=& -\pi_{\rho}-2\,(\lambda^+ +\lambda^-)^{-1}\psi\,[\,
-4\,(\dot{\rho}-\dot{\phi})\nn\\
& &+2\,(\lambda^+ -\lambda^-)(\rho '-\phi ')
+2\,(\lambda^+ -\lambda^-) '\,]~,\\
\pi_{f}^{i} &=& (\lambda^+ +\lambda^-)^{-1}\,[\,2\,\dot{f}_{i}-
 (\lambda^+ -\lambda^-)f_{i}'\,]~.
\nn\ee
So, $\pi_{\pm}^{\lambda}$  are primary
constraints, and the Dirac algorithm further
leads to the secondary constraints, the Virasoro constraints
\be
\varphi_{\pm} &=& -2\mu^2\exp 2(\rho-\phi)
+(\partial_{\sigma}-Y_{\pm})\,(\psi '\mp\frac{1}{2}\,\pi_{\rho})
+\frac{1}{4}\,\sum_{i=1}^{n}\,(\,\pi_{f}^{i}\pm f_{i} '\,)^2~,\\
\mbox{with}  & & Y_{\pm}\equiv (\rho-\phi) '
\pm\frac{1}{4\psi}\,(\pi_{\rho}+\pi_{\phi})~,
\ee
which satisfy under Poisson bracket the Virasoro algebra
\be
\{ \varphi_{\pm} ( \sigma )\ ,\ \varphi_{\pm} ( \sigma') \}_{\rm PB} &=&
\mp
(\ \varphi_{\pm} ( \sigma ) \partial_{\sigma '} - \varphi_{\pm} ( \sigma')
\partial_{\sigma}\ )\  \delta ( \sigma - \sigma')~, \\
\{ \varphi_{\pm} ( \sigma )\ ,\ \varphi_{\mp} ( \sigma')
\}_{\rm PB} &=& 0 \nn\ .
\ee
Since there are four first-class constraints,
$\pi_{\pm}^{\lambda}$ and $\varphi_{\pm}$, the theory without
the matter would have no physical degree of freedom.
\subsection{Extended phase space and BRST charge}
According to BFV \cite{bfv}, we define
the extended phase space by
including  to the classical phase space the
ghost-auxiliary field sector
\be
( {\cal C}^{\rm A}\ ,\ \overline{{\cal P}}_{\rm A} )\ ,\
( {\cal P}^{\rm A}\ ,\ \overline{{\cal C}}_
{\rm A} )\ ,\
( N^{\rm A}\ ,\ B_{\rm A} )\ ,
\ee
where  ${\rm A} ( = {\lambda}^{\pm},~\pm)$ labels the  first-class
constraints, $\pi_{\pm}^{\lambda}$ and $\varphi_{\pm}$.
$ {\cal C}^{\rm A}$ and ${\cal P}^{\rm A}$ are the BFV ghost fields
carrying one unite of ghost number,
$\mbox{gh}({\cal C}^{\rm A}) = \mbox{gh}({\cal P}^{\rm A}) =1$,
while $~~\mbox{gh}(\overline{{\cal P}}_{\rm A}) =
\mbox{gh}(\overline{{\cal C}}_ {\rm A}) = -1$
for their canonical momenta, $\overline{{\cal P}}_{\rm A}$ and
$\overline{{\cal C}}_ {\rm A}$.
The last canonical pairs in (17) are
auxiliary fields and carry no ghost number.
We assign $0$ to the canonical dimension of $
\phi, {\lambda}^{\pm}$ and $\rho$,
and correspondingly $+1$ to $\pi_{\phi},\pi^{\lambda}_{\pm}$, and
$\pi_{\rho}$. The canonical
dimensions of  ${\cal C}_{\lambda}^{\pm},~
\overline{{\cal P}}_{\pm}$
and $~\overline{{\cal P}}^{\lambda}_{\pm}$
are fixed only relative to that of ${\cal C}^{\pm}$. For
 $c~ \equiv \mbox{dim}({\cal C}^{\pm})$, we have:
\be
\mbox{dim}({\cal C}_{\lambda}^{\pm})~=~1+c~ ,~
\mbox{dim}(\overline{{\cal P}}_{\pm})~ =~ 1-c~
,~  \mbox{dim}(\overline{{\cal P}}^{\lambda}_{\pm})~ =~ -c~.
\ee

Given the first-class constraints with the corresponding algebra
(16), we can
construct a BRST charge \be
 Q &=& \int d \sigma [\ {\cal C}_{\lambda}^{+} {\pi}^{\lambda}_{+}
+ {\cal C}_{\lambda}^{-} {\pi}^{\lambda}_{-}
 + {\cal C}^{+} ( \varphi_{+} + \overline{{\cal P}}_{+}
{\cal C}^{+\prime} ) \nn \\
 & & + {\cal C}^{-} ( \varphi_{-} -\overline{{\cal P}}_{-}
{\cal C}^{-\prime} )
 + B_{\rm A} {\cal P}^{\rm A}\ ]~,\ (A=\lambda^{\pm},\pm)\\
& &\mbox{with gh}(Q) = 1 ,\mbox{dim}(Q)=1+c~,\nn
\ee
which satisfies under Poisson bracket the nilpotency condition
\be
\{Q~,~Q\}_{\rm PB} &=&0~.
\ee
At this stage, we would like to emphasize that the BRST charge (19) is given
prior to gauge fixing.
The BRST transformation of $A$ is then defined as
\be
\delta\,A &=& -\{Q~,~A\}_{\rm PB}~,
\ee
and one finds that
\begin{eqnarray}
\delta f^{i} & =& {1 \over 2} [\ {\cal C}^{+}(\pi_{f}^i + f_{i})
+ {\cal C}^{-} (\pi_{f}^i - f_{i})\ ] \ ,\nonumber \\
\delta \pi_{f}^i& =& {1 \over 2}
[\ {\cal C}^{+}(\pi_{f}^i + f_{i})
- {\cal C}^{-}(\pi_{f}^i - f_{i})\ ]^{\prime}  \ ,\nn\\
\delta \rho &=&{1 \over 2} [\ {\cal C}^{+ {\prime}} -{\cal C}^{-
{\prime}} +({\cal C}^{+} +{\cal C}^{-})\,\rho '+
({\cal C}^{+} +{\cal C}^{-})\frac{1}{4\psi}(2\pi_{\rho}+\pi_{\phi})~]
\ ,\nn\\
\delta \pi_{\rho} &=&4\mu^2(\,{\cal C}^{+} +{\cal C}^{-}\,)
\exp 2(\rho-\phi)-\frac{1}{2}\,[\,({\cal C}^{+} -
{\cal C}^{-})\,\pi_{\rho}\,]
^{\prime}-[\,({\cal C}^{+} +{\cal C}^{-})\,\psi '\,]^{\prime}~ ,\nn\\
\delta \phi &=&{1 \over 2} [\ ({\cal C}^{+} +{\cal C}^{-})\,\phi '+
({\cal C}^{+} +{\cal C}^{-})\frac{1}{4\psi}\,\pi_{\rho}~]
\ ,\nn\\
\delta \pi_{\phi} &=&-4\mu^2(\,{\cal C}^{+} +{\cal C}^{-}\,)
\exp 2(\rho-\phi)+\frac{1}{2}\,[\,({\cal C}^{+} -
{\cal C}^{-})\,\pi_{\phi}\,]
^{\prime}\nonumber\\
& &
+[\,({\cal C}^{+}
 +{\cal C}^{-})\,\psi '\,]^{\prime}~ +2\,[\,({\cal
C}^{+} +{\cal C}^{-})\, (\rho '-\phi ')\,]^{\prime}\,\psi+
2\,({\cal C}^{+\prime\prime} +{\cal C}^{-\prime\prime})\,\psi
\nonumber\\
& &
-({\cal C}^{+} +{\cal C}^{-})\frac{1}{4\psi}\,
(\,\pi_{\rho}^2+\pi_{\rho}\pi_{\phi}\,)~,\\
\delta {\cal C}^{\pm} & =& \pm
{\cal C}^{\pm}{\cal C}^{\pm\prime}\ ,~
\delta {\cal C}_{\lambda}^{\pm}\ =\ 0~,
{}~\delta {\lambda}^{\pm}~ =~ {\cal C}_{\lambda}^{\pm}~,\nn\\
\delta \overline{{\cal P}}_{\pm} & =&
- \Phi_{\pm}~=~-\varphi_{\pm} \mp
( \, 2 \overline{{\cal P}}_{\pm} \, {\cal C}^{\pm \prime}+
\overline{{\cal P}}_{\pm}^{\prime} \, {\cal C}^{\pm} \, )~ ,\nn\\
\delta\overline{{\cal
P}}^{\lambda}_{\pm} &=& -\pi^{\lambda}_{\pm}\ ,\  \delta N^{A}\ =\ {\cal
P}^{A}\ ,\ \delta\overline{C}_{A}\ =\ - B_{A}\ ,\nn\\
\delta{\pi}^{\lambda}_{\pm}
&=& \delta {\cal P}^{A}~=~ \delta B_{A}\ =\ 0~.\nn
\end{eqnarray}

\subsection{The fundamental brackets}
Gauge fixing appears in defining the total Hamiltonian $~H_{\rm T}$.
Since the canonical Hamiltonian
vanishes in the present case,
it is given by the BRST variation of gauge fermion ${\Psi}$, i.e.
\be
H_{\rm T}=\{ Q~,~\Psi \}_{\rm PB},
\ee
where the standard gauge fermion has the form
\be
\Psi &=&\int d \sigma [\ \overline{{\cal C}}_{\rm A} {\chi}^{\rm A} +
\overline{{\cal P}}_{\rm A} N^{\rm A}\ ]\ ,\ (A=\lambda^{\pm},~\pm)\ .
\ee
Eq. (23) immediately leads to
\be
\{Q\ ,\ H_{\rm T}\}_{\rm PB}\ &=&\frac{d}{dt}Q~ =~ 0\ .
\ee

In terms of the phase space variables,
the charge $W$, which generates the non-linear symmetry transformation (2),
can be written as
\be
W &=&-\frac{1}{2}
\int d\sigma\,(\pi_{\phi}+\pi_{\rho})\psi^{-1}~=~
-\int d\sigma\,(Y_{+}-Y_{-})~,
\ee
where $Y_{\pm}$ are defined in (15), and its
Poisson brackets with $Q$ and $H_{\rm T}$ are
\be
\{Q~,~W \}_{\rm PB} &=&0~, \\
\{H_{\rm T},~W\}_{\rm PB} &=&0~.
\ee
Eqs. (20), (25), (27) and (28) are the basic bracket relations which
exhibit the  2D general covariance and the global invariance at the classical
level.
 Note however that because of eq. (23),
the bracket relations (25) and (28)
are consequences of (20) and (27). Therefore,
only
(20) and (27) must be regarded as the fundamental bracket relations .

\subsection{Gauge-fixed action}
The BRST invariant master action of
the theory (1) is given by \cite{bfv}
\be
S_{\Psi} &=& \int d^{2}\sigma\ \{\
\sum_{i=1}^{n}\,\pi_{f}^{i}\dot{f}_{i}
+\pi^{\la}_{+}\dot{\la}^{+}+\pi^{\la}_{-}\dot{\la}^{-}
+\pi_{\phi}\dot{\phi}+\pi_{\rho}\dot{\rho}
+\overline{{\cal P}}_{A}\dot{{\cal C}}^{A} \nn\\
& &-{\cal H}^{\rm cl}-{\cal H}^{\rm gf}
-{\cal H}^{\rm FP}\ \}\ ,\ (A={\la}^{\pm},\pm)~,
\ee
where
\footnote{We have suppressed here the Legendre terms,
${\overline{{\cal C}}}_{A} {\dot{{\cal P}}}^{A}
+B_{A} \dot{N}^{A} = - \delta ({\overline{{\cal C}}}_{A} \dot{N}^{A})$, by
 shifting the gauge fermion, $\Psi \rightarrow \Psi
+ ({\overline{{\cal C}}}_{A} {\dot{N}}^{A})$.}
\be
{\cal H}^{\rm cl} &=&
\varphi_{+}\,N^{+} +\varphi_{-}\,N^{-}
+\pi^{\la}_{+}N^{+}_{\lambda}
+\pi^{\la}_{+}N^{+}_{\lambda}~,\nn\\
{\cal H}^{\rm gf} &=&B_{A}\chi^{A}~, \nn\\
{\cal H}^{\rm FP} &=&
 \overline{{\cal C}}_{A}\ \de\chi^{A}
+\overline{{\cal P}}_{A}{\cal P}^{A}
+ (\ 2\overline{{\cal P}}_{+}
{\cal C}^{+\prime} +\overline
{{\cal P}}_{+}^{\prime} {\cal C}^{+}\ )\ N^{+} \nn\\
& &-(\ 2 \overline{{\cal P}}_{-}{\cal C}^{-\prime} +
\overline {{\cal P}}_{-}^{\prime} {\cal C}^{-}\ )\ N^{-}~.\nn \ee

To illustrate how to proceed to obtain a gauge-fixed
action, let us discuss the conformal gauge-fixing for definiteness.
Terms in ${\cal H}^{\rm cl}$ suggest that $N^{\pm}$
play the role for $\lambda^{\pm} ~= ~
(\sqrt{-g}\pm g_{01})/g_{11}$.
So it is natural to impose $N^{\pm}=\lambda^{\pm}$. The conformal gauge
then corresponds to
\be
\chi_{\lambda}^{\pm} &=& \lambda^{\pm}-N^{\pm}~,~
\chi^{\pm} = N^{\pm}-1~.
\ee
Many variables such as $\lambda^{\pm},\pi^{\lambda}_{\pm},
N^{\pm},B_{\pm}, {\cal P}^{\pm},
\overline{{\cal P}}_{\pm}$, etc. become non-dynamical in the gauge,
and we can eliminate them
by means of the equations of motion like
\be
\overline{{\cal P}}_{\pm} &=&-\overline{{\cal C}}_{\pm}~.
\ee
In doing so, we arrive at the conformal-gauge action
\be
S_{\rm CG} &=& S_{\rm CG}^{\rm cl}
-\int d^2\sigma\,(\,\overline{{\cal C}}_{+}\,\partial_{-}{\cal C}^{+}
+\overline{{\cal C}}_{-}\,\partial_{+}{\cal C}^{-}\,)~,
\ee
where $S^{\rm cl}_{\rm CG}$ is given in (4).
The gauge-fixed form of the BRST charge can be similarly obtained from
the gauge-independent one (19):
\be
Q_{\rm CG} &=&Q_{\rm CG}^{(+)}+Q_{\rm CG}^{(-)}~,\nn\\
Q_{\rm CG}^{(\pm)}&=&\pm\int d\sigma^{\pm}\,{\cal C}^{\pm}\,\{\,
-\partial_{\pm}(\rho-\phi)\,\partial_{\pm}\psi+
\frac{1}{2}\partial_{\pm}\partial_{\pm}\psi \nn\\
& &+\frac{1}{4}\,\sum_{i=1}^{n}(\partial_{\pm}f_{i})^2
-\frac{1}{2}\overline{{\cal C}}_{\pm}\,\partial_{\pm}{\cal C}^{\pm}\,\}~.
\ee
In a similar way, one can find the gauge-fixed action in different
gauges. In section 6  , we will work in the light-cone gauge to
perform explicit calculations.

\section{Anomalies}
\subsection{Consistency conditions}
The ``nilpotency of $Q$" expressed in eq. (20) means that the
underlying  constraints in the theory are first-class while eq. (27)
exhibits the presence of the global symmetry. At the quantum level, $Q$ and
$W$ must be suitably regularized to become well-defined operators because
they are composite operators. We mean by an anomaly that the algebraic
structure which is to be traced back to a symmetry of a classical theory
can not be maintained upon quantization.
So, anomaly arises if $Q$ and $W$ do no satisfy
the fundamental algebra under commutator.
 In I, we have assumed that
anomalous Schwinger terms, which exhibit an
anomaly in the algebra, may be expanded in $\hbar$ as
\be
[Q\ ,\ Q] &= & i\sum_{m=2}\hbar^m\Omega^{(m)}~,
\ee
\be
[ Q\ ,\ W] &=& \frac{i}{2}\sum_{m=2}\hbar^m \Xi^{(m)}\ ,
\ee
($[\ ,\ ]$ denotes super-commutator), and that
the super-commutation relations
between $Q$ and $W$ satisfy:
\newline
(i) (anti-) symmetry,
\be
[A\ ,\ B]
&=&-(-1)^{\vert A\vert\vert B\vert}[ B\ ,\ A]\ ,\nn
\ee
(ii) bilinearity
\be
[A\ ,\ B+C]&=&[A\ ,\ B]+[A\ ,\ C]\ ,\nn
\ee
(vi) derivative property (super-Jacobi identity)
\be
& &[A\ ,\ [B\ ,\ C]\ ]
+(-1)^{\vert C\vert(\vert A\vert+\vert B\vert)}[C\ ,\ [A\ ,\ B]\ ]\nn \\
& &+(-1)^{\vert A\vert(\vert B\vert+\vert C\vert)}
[B\ ,\ [C\ ,\ A]\ ] \ =\  0~,
\ee
and (iv)
\be
[A~,~B]&=&i\hbar\,\{A~,~B\}_{\rm PB}+O(\hbar^2)~,
\ee
where $\vert A\vert$ is the grassmann parity of the operator $A$
and can be either even ($=0\ \mbox{mod}\ 2$) or odd
($=1\ \mbox{mod}\ 2$).
One easily observes that the outer commutators in the super-Jacobi
identities for $Q$ and $W$,
\be
[Q\ ,[Q\ ,\ Q]\ ]&=&0\ ,\\
2\ [Q\ ,\ [Q\ ,\ W]\ ]+[W\ ,\ [Q\ ,\ Q]\ ]&=&0\ ,
\ee
define a set of consistency conditions at each order in $\hbar$ in terms of
Poisson brackets.
In the lowest order, they are \cite{FIKA,fik}
\be
\delta\Omega^{(2)} &= &0~,\\
\delta\Xi^{(2)} &= &\{W\ ,\ \Omega^{(2)}\}_{\rm PB}\  .
\ee

The true anomalies $\Omega^{(2)}$ and $\Xi^{(2)}$ should be
cohomologically non-trivial.
To see this, we first observe that, if $\Omega^{(2)}$ and
$\Xi^{(2)}$ are solutions, then $\Omega^{(2)} + \delta X$
and  $\Xi^{(2)} +
\{ W\ ,\  X\}_{\rm PB} + \delta Y$ are solutions too.
It is then easy to find that $X$ and $Y$ can be
absorbed to order $\hbar^2$ into a re-definition of $Q$ and $W$,
defined by $Q\rightarrow  Q-(\hbar X/2)~\mbox{and}~  W\rightarrow
W - (\hbar Y/2)$.
If furthermore there is no non-trivial solution to (40)
(this will be the case in the CGHS theory),
the non-trivial $\Xi^{(2)}$ is a non-trivial solution to the homogeneous
part of (41):
\be
\delta\,\Xi^{(2)}_{h}&=&0~.
\ee

One can also convince oneself
that the consistency conditions
of $O(\hbar^3)$ are exactly the same as those of $O(\hbar^2)$ if
there is no non-trivial solution at $O(\hbar^2)$.
Therefore, the non-existence
of the non-trivial solutions to the consistency conditions (40) and (42)
ensures the non-existence
of those to the higher order consistency conditions.
This is the non-renormalization theorem \cite{adler} in our language.
\subsection{Solutions in the conformal gauge}
To find possible anomalies, we thus have to solve
the classical, algebraic
problem defined by (40) and (41) (or (42) if there is no non-trivial solution
to (40)). In I we have looked for solutions
$\Omega^{(2)}$ and $\Xi^{(2)}$
in the form  \be
\Omega^{(2)} &=& \int d \sigma \omega\ ,\\
\Xi^{(2)} ~(\Xi^{(2)}_{h})&=& \int d \sigma \xi~(\xi_{h})\ ,
\ee
where $\omega$  and $\xi~ (\xi_{h})$ are polynomials of local operators with
$~\mbox{gh}(\omega) = 2~$, $~\mbox{gh}(\xi) = \mbox{gh}(\xi_{h})=1~$,
$~\mbox{dim}(\omega) = 3~+~2c$, and $~\mbox{dim}(\xi) = \mbox{dim}(\xi_{h})
=2~+~c$.
We have first divided the total
phase space, according to the action of $\delta$,
into two sectors,
\be
& &{\rm S}_{1}\ \mbox{ consisting of}\
(f_{i}\ ,\ \pi_{f}^{i})\ ,(\rho\ ,\ \pi_{\rho})\ ,\ (\phi\ ,\ \pi_{\phi})\
 \mbox{and}\ ({\cal C}^{\pm}\ ,\ \overline{{\cal P}}_{\pm})\ ,  \nn\\
& &
{\rm S}_{2}\ \mbox{consisting of all the other fields}\ ,
\ee
such that the $\delta$ operation closes on each sector.
The ${\rm S}_{2}$-sector is BRST trivial because it is made of pairs
$(U^{a}\ ,\ V^{a})$
 with $\delta_{2} U^{a}= \pm V^{a}$.
As shown in ref. \cite{FIKA}, there exists no non-trivial solution to (40)
((42)) if $\omega\,(\xi_{h})$  contains the ${\rm S}_{2}$-variables.
We then have used
the linear independence of the generalized Virasoro constraints,
$\Phi_{\pm}\equiv-\delta\,\overline{{\cal P}}_{\pm}$,
and the fact that $\overline{{\cal P}}_{\pm}$ is the only one which produces
$\overline{{\cal P}}_{\pm}$  under $\delta$, to  conclude that
$\overline{{\cal P}}_{\pm}$ can not be involved in the non-trivial part
of $\omega$ and $\xi_{h}$.
Therefore, $\omega$ and $\xi_{h}$ are functions only
of $f_{i}, \pi_{f}^{i},\rho, \pi_{\rho} ,\phi , \pi_{\phi}$,
${\cal C}^{\pm}$ and their spatial derivatives.

The solutions to (40) and (41) are gauge-independent because all the
quantities involved there are defined prior to gauge-fixing.
To proceed, we have made the basic assumption in I
which is based on the time-independent reparametrization invariance.
Here we would like to present the conformal-gauge equivalent of the proof
that there are no non-trivial solutions to (40) and (42).

Using the equation of motion (5)
and those for the ghost fields in the conformal gauge, one easily
obtains the conformal
fields, e.g.,
\be
Y_{\pm} & =& \pm\partial_{\pm}(\rho-\phi)~,
{}~~(\mbox{dim}(Y_{\pm})~=1)~,\nn\\
F_{\pm}^{i}&=
&\partial_{\pm}f_{i}~,~~(\mbox{dim}(F_{\pm})~=1)~,\nn
\\ G_{\pm}&=
&\frac{1}{2}\,\partial_{\pm}\partial_{\pm}\psi
-\partial_{\pm}(\rho-\phi)\,\partial_{\pm}\psi ~,
{}~~(\mbox{dim}(G_{\pm})~=2)~,\\
{\cal C}^{\pm} & ,& (\mbox{dim}({\cal C}^{\pm})~=c)~.\nn
\ee
The BRST transformations of these quantities
are closed:
\be
\delta\,Y_{\pm} &=&\pm\frac{1}{2}\,\partial_{\pm}[\,(\frac{1}{2}
\partial_{\pm}\pm Y_{\pm})\,{\cal C}^{\pm}]~,\nn\\
\delta\, F_{\pm}^{i} &=&\frac{1}{2}\partial_{\pm}
(\,{\cal C}^{\pm}F^{i}_{\pm})~,\nn\\
\delta\,G_{\pm} &=&\frac{1}{2}{\cal C}^{\pm}\,\partial_{\pm}
G_{\pm}+\partial_{\pm}{\cal C}^{\pm}\,G_{\pm}~,\\
\delta\,{\cal C}^{\pm}&=&\frac{1}{2}{\cal C}^{\pm}
\,\partial_{\pm}{\cal C}^{\pm}~,\nn
\ee
The assumption of I is equivalent to that
the non-trivial solutions to (40) and (42) are functions only of the
conformal fields listed above
\footnote{$\ol{\cal C}_{\pm}$, which are $-\ol{\cal P}_{\pm}$
in the conformal gauge (see eq. (31), are also conformal fields,
but they are not involved in the non-trivial solutions as
we have argued above.}
and that they respect the discrete symmetry
defined by  ${\cal C}^{\pm} \rightarrow {\cal C}^{\mp}, \partial_{\sigma}
\rightarrow - \partial_{\sigma}$.

We first consider the $Q^2$-anomaly in the conformal gauge.
In that gauge, the BRST charge is
split into its chiral components, $Q_{\rm CG}^{(+)}$ and
$Q_{\rm CG}^{(-)}$
(see eq. (33)), which simplifies the analyses because
we may further assume \footnote{As a matter of fact, (48) is not
an assumption; we justify it in appendix.}
that \be
[\,Q_{\rm CG}^{(+)}~,~Q_{\rm CG}^{(-)}\,]&=&0~.
\ee
It therefore follows
\be
\Omega_{\rm CG}^{(m)} &=&\Omega_{\rm CG}^{(+m)}+\Omega_{\rm
LG}^{(-m)}~,~(m=2,\cdots)~,
\ee
where
\be
[\,Q_{\rm CG}^{(\pm)}~,~Q_{\rm CG}^{(\pm)}\,]&=&
i\sum_{m=2}\,\hbar^{m}\,\Omega_{\rm CG}^{(\pm m)}~=~
i\sum_{m=2}\,\hbar^{m}\,\int d\sigma \,\omega^{(\pm m)}_{\rm CG}~.
\ee
With this in mind,
we write the most general form for $\omega_{\rm CG}^{(+2)}$:
\be
\omega_{\rm CG}^{(+2)} &=&{\cal C}^+ \,\partial_{+}{\cal C}^{+}\,
\{ ~\kappa_{1}\,G_{+}+\kappa_{2}\,(Y_{+})^2+
\kappa_{3}\,\partial_{+} Y_{+}+
\sum_{i=1}^{n}\,\kappa_{4}^{(i)}\,(F_+^i)^2~\}\nn\\
& &+\kappa_{\rm KO}\,\partial_{+} {\cal C}^{+} \,(\partial_{+})^{2}
\,{\cal C}^{+} ~,
\ee
where $\kappa$'s are constants independent of the fields.
 (For $\omega_{\rm
CG}^{(-2)}$, we have to do the same analysis.)
By requiring that $\delta\omega_{\rm CG}^{(+2)}$ be at most a
total derivative, we find the conditions on $\kappa$'s
\be
\kappa_{2} &=& -\kappa_{3}~.
\ee
But if eq. (52) is satisfied,
$\omega_{\rm CG}^{(+2)}$ is a trivial term because
\be
\omega_{\rm CG}^{(+2)} &=& -2\delta\,[~\kappa_{1}\,
{\cal C}^+ \,G_{+}
+\kappa_{2}\,{\cal C}^+(Y_{+})^2~
+\sum_{i=1}^{n}\kappa_{4}^{(i)}\,{\cal C}^+\,(F_{+}^{i})^2
+2 \kappa_{\rm KO}\,\partial_{+}{\cal C}^{+}\,Y_{+}\,] \nn\\
& &
+\partial_{+}\,[\,(\,-\kappa_{2} +2\kappa_{\rm KO}\,)\,Y_{+}\,
{\cal C}^{+}\,\partial_{+} {\cal C}^{+} ~  ]~,
\ee
implying that there is no non-trivial solution to (40) to $O(\hbar^2 )$
and hence
to any finite order of $\hbar$. Note that the Kato-Ogawa anomaly
term \cite{KOA}, the last term on the r.h.s. of eq. (51), is trivial.
\footnote{That is, the central extension of the Virasoro
algebra is cohomologically trivial. This is our
language to the observation of refs. \cite{cham,terao1}.}
As for anomaly in the non-linear global
symmetry, we would like  to slightly modify the treatment of I. This is
because the global symmetry appears as a residual, local symmetry  in the
conformal gauge: One can easily verify that the gauge-fixed action, $S_{\rm
CG}$ given in (32), is invariant under the bosonic transformation
\be
\delta_{\rm B}^{(\pm)}\,\rho
&=&\delta_{\rm B}^{(\pm)}\,\phi~=~-\frac{1}{2}\zeta_{\pm}\,
\exp 2\phi~,
\ee
as well as under the fermionic transformation
\be
\delta_{\rm F}^{(\pm)}\,\rho &=&\delta_{\rm F}^{(\pm)}\,\phi~=~
\mp\frac{1}{4}\,\partial_{\pm}\eta_{\pm}\,{\cal C}^{\pm}\,
\exp2\phi~,\nn\\
\delta_{\rm F}^{(\pm)}\,\overline{{\cal C}}_{\pm} &=&
\pm\frac{1}{2}(\partial_{\pm})^2 \eta_{\pm}
\mp\partial_{\pm}\eta_{\pm}\,\partial_{\pm}(\rho-\phi)~,\\
\delta_{\rm F}^{(\pm)}\,{\cal C}^{\pm} &=&0~, \nn
\ee
where $\eta_{+}\,(\eta_{-})$ is a fermionic function of $\sigma^{+}\,
(\sigma^{-})$ with the ghost number $-1$
while $\zeta_{+}\,(\zeta_{-})$ is a bosonic function of $\sigma^{+}\,
(\sigma^{-})$. The generators for the  bosonic and
fermionic transformations are
\be
W_{\rm B}^{(\pm)}[\zeta] &=&
\mp\int d\sigma^{\pm}\,\zeta_{\pm}\,\partial_{\pm}(\rho-\phi)~,\\
W_{\rm F}^{(\pm)}[\eta] &=&\pm\int d\sigma^{\pm}\,\partial_{\pm}\eta_{\pm}\,
[\frac{1}{2}\partial_{\pm}+\partial_{\pm}(\rho-\phi)]\,{\cal C}^{\pm}~.
\ee
Together with $Q_{\rm CG}^{(\pm)}$ they form a closed algebra
under Poisson bracket:
\be
\{Q_{\rm CG}^{(\pm)}~,~W_{\rm B}^{(\pm)}[\zeta]\}_{\rm PB} &=&
W_{\rm F}^{(\pm)}[\zeta]~,
\ee
where $W$'s commute with each other.
Since the original non-linear transformation (2)
corresponds to the bosonic one (56) with $\zeta_ +=\zeta_- =\Lambda/2$,
the closure of the
algebra (58) under commutator
implies the absence of anomaly in the non-linear
symmetry (2) automatically.

Appendix is devoted to give
an algebraic proof for its absence. So our main conclusion
in this section is that
there are no non-trivial solutions to the consistency conditions
(40) and (42) to any finite order in $\hbar$.

\section{Matter contribution to anomalies}
\subsection{Basic idea}
In the previous section, we have performed an algebraic
analysis on possible anomalies, which is based on the consistency
conditions (40) and (42). They have been derived from the basic assumptions
(i)-(v). Of those, the Jacobi identity
(36) is undoubtedly the most crucial one. Although there are no
examples of Jacobi-identity violating commutator known in
two dimensions (at least to our knowledge),
there exist some examples in four dimension  \cite{jacobi} .
So, it is certainly desirable to explicitly show
the absence of anomalies in
CGHS theory.

So far, the single, global symmetry defined in (2)
has been made responsible for two independent,
physical consequences; degeneracy of the black hole back ground (7)
and the fact that the conformal factor of $g_{\al\bt}$ does not
couple to the matter fields. It is more
convenient
if we can treat these two things independently and individually.
To this end,
we consider an action similar to $S^{\rm cl}$ given in (1)
but with the matter coupling replaced by
\be
 -\frac{1}{2}\,\sqrt{-\overline{g}}\,\sum_{i=1}^{n}
\,\overline{g}^{\alpha\beta}
\ \partial_{\alpha}f_{i}\,\partial_{\beta}f_{i}~,
\ee
where $\ol{g}_{\al\bt}$ is supposed to differ from $g_{\al\bt}$
only in the conformal factor.
We parametrize $\overline{g}_{\alpha\beta}$ in a similar way as we
did for $g_{\alpha\beta}$ in (11):
\be
\overline{g}_{\alpha\beta}&\equiv&
\left( \begin{array}{cc} -\lambda^{+}\lambda^{-} & (\lambda^{+}
-\lambda^{-})/2 \\
(\lambda^{+}-\lambda^{-})/2 & 1 \end{array} \right)
\exp 2\,\overline{\rho} ~.
\ee
Because of the local Weyl invariance of the matter coupling (59),
$\overline{\rho}$ does not show up in the action
at the classical level, and so the term (59)
transforms like a density under a two-dimensional general coordinate
transformation even if $g_{\al\bt}$ is regarded as the metric.
Therefore, two theories are equivalent at the classical level.
In this way, we have introduced two independent Weyl symmetries, global
and local, and
achieved to separate the theoretical origin
for degeneracy of the black hole back ground
from that for decoupling of the matter from
the conformal factor of $g_{\al\bt}$. If the local Weyl
symmetry turns out to be good at the quantum level, we may impose
different gauge conditions. In the $\ol{\rho}=\rho$-gauge,
for instance, we recover the original theory with
$g_{al\bt}$ in the matter coupling, and we can manifestly see
the decoupling of conformal factor of $g_{\al\bt}$ from the matter
in the $\ol{\rho}$-gauge.
\subsection{BRST symmetrization of the local  Weyl symmetry}
We denote the classical action with the matter coupling (59)
by $\overline{S}^{\rm cl}$, and follow the BFV procedure
as we have done in section 3. Every thing is the same, except
for the fact that there is an additional first-class constraint
\be
\pi_{\ol{\rho}} &\sim&0~,
\ee
where $\pi_{\ol{\rho}}$ is the conjugate momentum of $\ol{\rho}$.
Therefore, the BRST charge can be easily obtained from $Q$
(given in (19)):
\be
\ol{Q} &=&Q +\int d\sigma\,(~\frac{1}{2}\,{\cal C}^{\ol{\rho}}\,
\pi_{\ol{\rho}} +B_{\ol{\rho}}\,{\cal P}^{\ol{\rho}}~)~.
\ee

To determine the matter contribution to the anomalous Schwinger term
$\Omega$,
we will use the results of refs. \cite{PolA} and \cite{fujikawa} on the
path-integration of the matter fields $f_{i}$. Their results are
written in a covariant form, and so we first have to covariantize our
canonical formulation. (Since  $\ol{g}_{\al\bt}$
is supposed to couple to the matter, the covariantization has to
be made with respect $\ol{g}_{\al\bt}$ of course.)
This requires
elimination of various phase space variables by means
of the equations of motion, and for that we have to specify a gauge.
  In the standard form of the gauge fermion defined in eq. (24),
we have five gauge
conditions $\chi^{\rm A}$ with $A=\lambda^{\pm}\,,\,\pm\,,\,\ol{\rho}$.
To identify
 $g_{\al\bt}$ and $\ol{g}_{\al\bt}$
as well as the reparametrization ghosts and the
Weyl ghost with the extended-phase-space variables,
 we use two of them to impose the geometrization conditions
\cite{FIKM}
\begin{equation}
{\chi}_{\lambda}^{\pm}=\lambda^{\pm}-N^{\pm}~,
\end{equation}
while making  an (inessential) assumption that
$\chi^{\pm}$ and $\chi^{\ol{\rho}}$ do not contain
$${\pi}^{\lambda}_{\pm}~,~
\pi_{\ol{\rho}}~,~ \overline{{\cal C}}^{\lambda}_{\pm}~,~
\overline{{\cal P}}_{\pm}~,~
N_{\lambda}^{\pm},N^{\ol{\rho}}~,~\overline{{\cal P}}^{\lambda}_{\pm}~,~
\mbox{and}~ B^{\lambda}_{\pm}~.$$
One finds that
\begin{eqnarray}
\lambda^{\pm}&=&N^{\pm}~,~
N_{\lambda}^{\pm} = ~ \dot\lambda^{\pm}~,~  N^{\ol{\rho}} ~=
{}~ 2\dot{\ol{\rho}}~,\nonumber\\
{\cal P}^{\pm}&=&{\cal C}_{\lambda}^{\pm} ~
=~ \dot{\cal C}^{\pm}\pm{\cal C}^{\pm}\,N^{\pm\prime}
\mp{\cal C}^{\pm\prime}\,N^{\pm}
\end{eqnarray}
can be still unambiguously derived. Then one can verify
that the covariant
variables defined by
\begin{eqnarray}
C^{0}&\equiv&{\cal C}^{0}/N^{0}~,~C^{1}={\cal C}^{1}-
N^{1}{\cal C}^{0}/N^{0}~,\nn\\
C_W &\equiv&{\cal C}^{\ol{\rho}}-C^{0}N^{\ol{\rho}}-
2\,C^{1}\,{\ol{\rho}}^{\prime}
-2\,C^{0\prime}\,N^{1}-2\,C^{1\prime}~,\nn\\
g_{\alpha\beta}&\equiv&
\left( \begin{array}{cc} -N^{+}N^{-} & (N^{+}
-N^{-})/2 \\
(N^{+}-N^{-})/2 & 1 \end{array} \right)\exp 2\rho~,\\
\ol{g}_{\al\bt}&=&g_{\al\bt}\,\exp\,2\,(\ol{\rho}-\rho)~,\nn
\end{eqnarray}
obey the covariant BRST transformation rules \cite{fujikawa}
\begin{eqnarray}
\delta \ol{g}_{\alpha\beta}&=&C^{\gamma}\, \partial_{\gamma}
\ol{g}_{\alpha\beta}
+\partial_\alpha C^{\gamma}\, \ol{g}_{\gamma\beta}
+\partial_\beta C^{\gamma}\, \ol{g}_{\alpha\gamma}+C_W\,
\ol{g}_{\alpha\beta}~,\nn\\
\delta g_{\alpha\beta}&=&C^{\gamma}\, \partial_{\gamma} g_{\alpha\beta}
+\partial_\alpha C^{\gamma}\, g_{\gamma\beta}
+\partial_\beta C^{\gamma}\, g_{\alpha\gamma}~,\\
\delta C^{\alpha} &=&C^{\gamma}\, \partial_{\gamma} C^{\alpha}~,~
\delta C_{W} =C^{\gamma}\, \partial_{\gamma} C_{W}~,\nn
\end{eqnarray}
where $ C^{\pm} = C^{0} \pm C^{1}$
and similarly for other quantities. Not that the $\tau $-derivatives in
eq. (66) are  exactly those which appear in
eq. (64).

\subsection{Triviality of the matter contribution}
We are now in position to apply the results of refs.
\cite{PolA} and \cite{fujikawa}, and
find that the matter contribution to the anomalous Schwinger term
in $[\,\ol{Q}~,~\ol{Q}\,]$ is given by
(see also ref. \cite{man,IKS})
\be
i\,\Omega_{\rm m}&=&  i\,\kappa\int d\sigma\
\sqrt{-\ol{g}}\,[\
 C^{0}\ C_{W}\ R(\ol{g})+
\ol{g}^{0\beta}\ C_{W}\partial_{\beta}C_{W} +MC^{0}C_{W}~]~,\\
\kappa &=&-\frac{n}{24\pi}~,\nn
 \ee
where $n$ is the number of the matter fields.
In string theory, the $\Omega_{\rm M}$ is BRST non-trivial (as well known),
 but it is not the case here, because
\be
\Omega_{\rm m} &=&\Omega_{\rm KO}-\kappa\,\delta\,\int
d\sigma\,\vartheta~,\\ \Omega_{\rm KO}&=&-\kappa\,\int d\sigma
[\ ({\cal C}^{+\prime}  {\cal C}^{+\prime\prime}) -
(+ \rightarrow - )\ ] \nn\\
&=&\kappa\,\delta\,\int d\sigma [~{\cal C}^{+\prime}\,
Y_{+}  +
(+ \rightarrow - )~]~,
\ee
where $\Omega_{\rm KO} $ is the Kato-Ogawa anomaly term \cite{KOA},
$Y_{\pm}$ is given in (15), and $\vartheta$ is given by
\cite{FIKM}
\begin{eqnarray}
\vartheta &=& \frac{1}{4}\ G_{+}^2 {\cal C}^{+} + G_{+}\,{\cal C}^{+\prime}
+\frac{1}{4}\ G_{-}^2 {\cal C}^{-} + G_{-}\,{\cal C}^{-\prime} \nn\\
& &
- \frac{1}{2} (G_{+} - G_{-})\, {\cal C}^{\ol{\rho}}
+\frac{M}{2}\,(\,{\cal C}^{+}+{\cal C}^{-})\,\exp 2\ol{\rho}\ ,\\
 G_{\pm} &=&  \frac{1}{{N}^{0}}~[~\pm  N^{\ol{\rho}} + 2({N}^{0} \mp
{N}^{1})~
\ol{\rho}^{\prime} \mp 2~ {N}^{1 \prime}~]\ .\nn
\end{eqnarray}
Therefore, the matter contribution can be completely absorbed into
a re-definition of $\ol{Q}$ so that its nilpotency can be maintained
\footnote{A similar observation has been made
for the Jackiw-Teitelboim model in refs. \cite{cham,terao1}.}.
This implies that the local Weyl symmetry of the matter coupling
(59) is quantum theoretically intact. (
The dilaton sector can not contribute to the anomaly
in that local Weyl symmetry.) So we explicitly have shown
that two theories, one with only $g_{\al\bt}$ and the other
with $\ol{g}_{\al\bt}$ in the matter coupling, are
equivalent. In other words, the quantum theory
corresponding to the action
$\ol{S}^{\rm cl}$  is gauge-equivalent to
that corresponding to $S^{\rm cl}$, and the decoupling
of the conformal factor of $g_{\al\bt}$ from the matter is manifest.

\section{Light-cone gauge fixing}
\subsection{Gauge-fixed form of $Q$ and $W$}
It is clear that the matter does not contribute to the anomaly
in the global Weyl symmetry because the path-integration of the matter
fields are supposed to be covariantized with respect $\ol{g}_{\al\bt}$.
This fact, together with the results
we have obtained in the previous section, leads to the observation that
the matter fields can never contribute non-trivially
to the anomalies we are
concerned with.

Therefore, we may completely switch off the matter coupling
to compute the possible anomalies. So we consider the original CGHS
theory without the matter fields in the modified light-cone gauge,
in which the theory takes the most simple form, and calculate
the anomalies in that gauge. It is the light-cone gauge for the
scaled metric $\hat{g}_{\al\bt}$ defined in eq. (10),
as proposed in ref. \cite{terao}.
\footnote{Some of the results we will obtain below have already  been found
in ref. \cite{terao}, but we would like to  present our results in
some detail for completeness.}

The modified light-cone gauge,
\be
\hat{g}_{+-} &=&\frac{1}{4}\,(\,\hat{g}_{00}-\hat{g}_{11}\,)~=~ -1/2~,\nn\\
\hat{g}_{--}
&=&\frac{1}{4}\,(\,\hat{g}_{00}+\hat{g}_{11}-2\,\hat{g}_{01}\,)~=~ 0~, \ee
is realized
by the gauge-fixing functions (in the gauge fermion $\Psi$)
\begin{eqnarray}
\chi^{+} &=& N^{+}-1~, \quad
\chi^{-} = [\,(N^{-}+1)\exp 2\,(\rho-\phi)\,]~ - 2~,
\end{eqnarray}
along with those given in (63).

To obtain the gauge-fixed action,
 we substitute the gauge fermion
(24) with $\chi$'s given in (63) and (72) into the
extended-phase-space action
(29), and then integrate out all the non-dynamical fields
by using the equations of motion as we did to obtain
the conformal gauge action (32); for instance,
\begin{eqnarray}
\pi_{\phi}&=&2\,\partial_{-}(\psi\exp 2(\rho-\phi))
-4\,\psi\rho' \nn\\
& &-2\,(\,{\cal C}^{+}\,\ol{\cal C}_{+}+
{\cal C}^{-}\,\ol{\cal C}_{-}\,)\exp 2(\rho-\phi)~,\nn\\
\pi_{\rho}&=&-2\,(\partial_{-}\psi)\,\exp 2(\rho-\phi)
+4\,\psi\phi'~ \\
& &+2\,(\,{\cal C}^{+}\,\ol{\cal C}_{+}+
{\cal C}^{-}\,\ol{\cal C}_{-}\,)\exp 2(\rho-\phi)~,\nn\\
 \overline{{\cal P}}_{+} &=&~
-\overline{{\cal C}}_{+}~,~\overline{{\cal P}}_{-}~=~-\exp 2\,(\rho-\phi)
\,\overline{{\cal C}}_{-}~.\nn
\end{eqnarray}
 The gauge-fixed action is then given by
\be
S_{\rm LG} &=&S_{\rm LG}^{\rm cl}+S_{\rm LG}^{\rm gh}~,\\
S_{\rm LG}^{\rm cl}&=&\int d^2 \sigma\,[\,4\,\mu^2-\partial_- \psi\,
\partial_- (\exp 2\,(\rho-\phi))\,]~,~(\psi=\exp -2\phi)~,\nn\\
S_{\rm LG}^{\rm gh}&=& -\,\int d^2 \sigma~
[~b_{++}\, \partial_{-} c^{+} + b\, \partial_{-} c_{+}~]~,
\nn\ee
where we have re-defined the ghost fields as \cite{fikt}
\begin{eqnarray}
c^{+} &=& {\cal C}^{+} ~,~
b~=~\overline{{\cal C}}_-~,\nn\\
c_{+} &=&{\cal C}^{-} +
(\,\exp 2\,(\rho-\phi)-1\,)~({\cal C}^{+} + {\cal C}^{-}) +
 \frac{\sigma^{-}}{2}\partial_{+}{\cal C}^{+}~,\\
b_{++} &=& \overline{{\cal C}}_+ -
(\,\exp 2\,(\rho-\phi)-1\,)\overline{{\cal C}}_{-}
+\frac{\sigma^{-}}{2}\partial_+ \overline{{\cal C}}_{-}~,\nn
\ee
in order to simplify the ghost sector. The re-defined
ghost fields satisfy the free equations of motion
\be
\partial_{-}\,c^{+} &=&\partial_{-}\,c_{+}~=~0,\nn\\
\partial_{-}\,b_{++}&=&\partial_{-}\,b~=0~.
\ee

Next we derive the gauge-fixed form of
the generator for the non-linear transformation (26),
$W_{\rm LG}$, and that of the BRST charge,
$Q_{\rm LG}$, by using
equations of motions (73) and (75), and
\be
(\partial_{-})^2 \psi &=&(\partial_{-})^2 \,\exp 2\,(\rho-\phi)
{}~=~0~.
\ee
Suppressing the total spatial derivative term,
we can easily obtain
\be
W_{\rm LG} &=&-\int d\sigma\,\partial_{-}(\exp 2(\rho-\phi))~.
\ee
After some
algebraic calculations, we also find that $Q_{\rm LG}$ takes the form
(see also refs. \cite{terao,fikt})
\be
Q_{\rm LG} &=&\int d\sigma\,\{\,c^+  (\,T_{++}^{\rm d}+
\frac{1}{2}\,T_{b++}+T_{b}\,)
+c_{+}\,T \,\}~,\\
T_{++}^{\rm d} &=&\frac{1}{2}\,\{ ~(\partial_+ )^2 \psi-
2\,\partial_- \partial_+ \psi+ 2\,\exp 2(\rho-\phi)\,
\partial_- \partial_+ \psi \nn\\
& &+\partial_{+}(\exp 2(\rho-\phi))\,
\partial_- \psi -\partial_{-}(\exp 2(\rho-\phi))\,
\partial_+ \psi +\sigma^{-}\partial_{+} T~ \} ~,\nn\\
T_{b++}& =& -\frac{1}{2}{\partial}_{+} b_{++}\,
c^{+} - \,b_{++}\, {\partial}_{+} c^{+}~,~ T_{b}~=~
{\partial}_{+} b\, c_{+}~,\nn\\
T&=&\frac{1}{2}\,\{ ~-4\,\mu^2-\partial_+ \partial_- \,\psi
-\partial_- (\exp 2(\rho-\phi)\,\partial_- \psi~\}~.\nn
\ee
We further use the equations of motion (77) to expand
$\psi$ and $\exp 2(\rho-\phi)$ according to
\be
\psi &=& \exp -2\phi~=~
\xi(\sigma^{+})+\frac{\sigma^{-}}{2}\,\xi_{-}(\sigma^{+})~,\nn\\
\exp 2(\rho-\phi) &=&
h_{++}(\sigma^{+}) +\frac{\sigma^{-}}{2}\,h_{+}(\sigma^{+})~.
\ee
In terms of $\xi$'s and $h$'s, $T_{++}^{\rm d}$ and $T$
 can be written as
\footnote{See also ref. \cite{terao}.}
\be
T_{++}^{\rm d}&=&\frac{1}{2}\,\{~(\partial_{+})^2\xi
-2\,\partial_{+}\xi_{-}+2\,h_{++}\,\partial_{+}\xi_{-}
+\partial_{+}h_{++}\,\xi_- -h_{+}\,\partial_{+}\xi~\}~,\\
T&=& \frac{1}{2}\,\{-4\,\mu^2 -\partial_{+}\xi_{-}-h_{+}\xi_{-}~\}~,\nn
\ee
and also $W_{\rm LG}$ becomes
\be
W_{\rm LG} &=&-\int d\sigma^{+}\,h_{+}(\sigma^{+})~.
\ee

We now have the facility to explicitly compute the anomalous
Schwinger terms $\Omega$ and
$ \Xi$ defined in eqs. (34) and (35).
\subsection{Explicit calculation of $\Omega$ and $\Xi$}
Until now, we have worked on the basis of the Poisson bracket.
To compute the anomalous Schwinger terms, we of course have to go to
the commutator and specify the ordering of operators.
To this end, let us first derive the commutation relations among
$\xi$'s and $h$'s (defined in (80)) from their original
Poisson brackets. Using eq. (73), one finds that
\be
\{\,\exp 2\,(\rho-\phi)(\tau,\sigma)~,
{}~(\partial_{-}\psi)(\tau,\sigma^{\prime})\,\}_{\rm PB}&=&
-\delta (\sigma-\sigma^{\prime})~,\nn\\
\{\,\psi(\tau,\sigma)~,
{}~[\partial_{-}\exp 2\,(\rho-\phi)](\tau,\sigma^{\prime})\,\}_{\rm PB}&=&
-\delta (\sigma-\sigma^{\prime})~,\\
\{\,\exp 2\,(\rho-\phi)(\tau,\sigma)~,
{}~\psi(\tau,\sigma^{\prime})\,\}_{\rm PB}&=&0~,\nn\\
\{\,[\partial_{-}\exp 2\,(\rho-\phi)](\tau,\sigma)~,
{}~(\partial_{-}\psi)(\tau,\sigma^{\prime})\,\}_{\rm PB}&=&0~,\nn
\ee
from which we deduce that
\be
[h_{++}(\sigma^{+})~,~\xi_{-}(\sigma^{\prime+})]&=&
[\xi(\sigma^{+})~,~h_{+}(\sigma^{\prime+})]~=~
-i\hbar\,\delta (\sigma^{+}-\sigma^{\prime+})~,
\ee
and other commutators vanish identically.

We define the positive and negative frequency parts of $J(\sigma^{+})
{}~(=h_{++}(\sigma^{+})~\cdots)$
by  \footnote{The step function is given by
$\theta (\pm\omega)=\int dx\,\delta ^{(\mp)}(x) e^{i\omega x}$.}
\be
J^{(\pm)} (\sigma^{+})& =&\int d\sigma^{\prime+}\,
{\delta}^{(\mp)}(\sigma^{\prime+}-\sigma^{+})\,
J(\sigma^{\prime+})~, \\
 {\delta}^{(\pm)}(\sigma^{+}) &=&
 \frac{\pm i}{2\pi(\sigma^{+} \pm i0+)}~,\nn
\ee
and employ the normal ordering prescription
with respect to $J^{(\pm)}$ to define the products
of operators ($J^{(+)}$ stands to the right of $J^{(-)}$).
We then split the commutation relation (84) into
\be
[h_{++}^{(+)}(\sigma^{+})~,~\xi_{-}^{(-)}(\sigma^{\prime+})]&=&
-i\hbar\,\delta^{(+)}
(\sigma^{+}-\sigma^{\prime+})~,\nn
\ee
\be
[h_{++}^{(-)}(\sigma^{+})~,~\xi_{-}^{(+)}(\sigma^{\prime+})]&=&
-i\hbar\,\delta^{(-)}
(\sigma^{+}-\sigma^{\prime+})~,
\ee
and similarly for $\xi$ and $h_{+}$, and also for $c^+ ~,~c_+ ~\cdots$.

As for the $Q^2$-anomaly, we can verify the result of ref. \cite{terao}
that $Q_{\rm LG}$ given in (79) with (81) satisfies the nilpotency condition.
The main reason of this is that the central charge, $c^{\rm d}$,
for the algebra
\begin{eqnarray}
[~T_{++}^{\rm d}(\sigma^{+})~,~T_{++}^{\rm d}(\sigma^{\prime+})~] &=&
i\hbar\,(\, T_{++}^{\rm d}(\sigma^{+}) +
T_{++}^{\rm d}(\sigma^{\prime+})
\,) \delta^\prime (\sigma^{+}-\sigma^{\prime+})\nonumber\\
& & - i\hbar^{2}\, \frac{c^{\rm d}}{24\pi}~
\delta^{\prime\prime\prime} (\sigma^{+}-\sigma^{\prime+})\
\end{eqnarray}
is exactly $+28$ \cite{terao} which is canceled by the ghost contribution
\cite{kpz,kuramoto}. ($T_{++}^{\rm d}$ is normal-ordered.)

The second commutator, $[\,Q_{\rm LG}~,~W_{\rm LG}\,]$,
can be trivially calculated because $W_{\rm LG}$ depends
on $h_{+}$ only linearly; the commutator vanishes identically.
 So $\Omega$ and $\Xi$ are zero in the (modified) light-cone
gauge, as we have expected from the algebraic consideration
in section 4.

Triviality of the contribution of the conformal matter
can be seen in the modified light-cone gauge too. Suppose
that the matter contribution to the energy-momentum tensor
is denoted by
$T^{\rm m}_{++}$ in that gauge, which satisfies the
Virasoro algebra (87) with the central charge $c^{\rm m}$.
The introduction of the matter modifies
the BRST charge (79) to
\be
Q^{\rm m}_{\rm LG} &=&Q_{\rm LG}+
\int d\sigma^+ \,c^+ \,T^{\rm m}_{++}~,
\ee
which is no longer nilpotent, i.e.,
\be
[\,Q^{\rm m}_{\rm LG}~,~Q^{\rm m}_{\rm LG}\,]&=&
\frac{i\hbar^2}{24\pi}(\frac{1}{8})\,c^{\rm m}\,
\int d\sigma^+\,\partial_{+}c^+ \,(\partial_{+})^2 c^+~,
\ee
where all the operator products are normal-ordered according to
(85).
But the r.h.s. of (89) is BRST trivial, because it can be
written as
\be
-\frac{i\hbar^2}{12\pi}(\frac{1}{8})\,c^{\rm m}\,
\delta\,\int d\sigma^{+}\,\partial_{+}c^+
\partial_{+}\,h_{+}~,
\ee
where $h_{+}$ is defined in (80).
Therefore, we can absorb this trivial anomaly into
a re-definition of $Q^{\rm m}_{\rm LG}$, and one finds
that the re-defined one is given by
\be
Q^{\rm m \prime}_{\rm LG}&=&Q^{\rm m}_{\rm LG}
-\frac{\hbar}{24\pi}(\frac{1}{8})\,c^{\rm m}\,
\int d\sigma^+ \partial_{+}c^+\,h_{+}~.
\ee
One can easily verify that $Q^{\rm m \prime}_{\rm LG}$
given above exactly
satisfies the quantum-nilpotency condition.

Before we close this section, we summarize its content.
We considered the CGHS action with the matter coupling
switched off, because we observed that
the matter fields $f_{i}$ can not non-trivially contribute
to the anomalous Schwinger terms $\Omega$ and $\Xi$.
Then we went to the (modified) light-cone gauge in which the theory
takes the most simple form so that we have to deal with
only free fields. The BRST charge and $W$ were expressed in terms of
those free fields. Having defined the normal ordering
prescription, we computed the relevant commutators and found
that $\Omega$ and $\Xi$ vanish identically in the
(modified) light-cone gauge.

The general result of ref. \cite{C} on the relation
between the trace anomaly
and the Hawking radiation in two-dimensions
remains of course correct. But our analysis suggests that one should first
 investigate whether Weyl anomalies are cohomologically trivial or not
in a given model before one applies the general result.
\section{Discussion}
There are some shortcomings in our investigation on anomalies
in the cosmological model of CGHS.
We find the following two may be serious:
\newline
(i)
We have not explicitly taken into account the black hole background
in computing anomalies. That is, we have assumed that
the invariance property of the theory is not influenced by
a space-time singularity. This assumption is certainly
a strong assumption, and should be carefully checked in future.
\newline
(ii)
We have neglected the infrared problem
in two-dimensions \cite{w}
(see also ref. \cite{na} and
references therein). If this problem is real, we have to
introduce an infrared cut-off which might violate the Weyl
invariance. In this connection, it may be
interesting to introduce massive matter fields and then
the mass-zero limit.
Some non-trivial interplay between
 the mass of the matter
and the infrared cut-off in the limit could be observed.
The  message of this paper is therefore that Weyl invariance can be
anomaly-free in two-dimensional dilaton gravity models and
the scale of its violation is presumably related to
the rate of the Hawking radiation in
two-dimensions.

It might be possible that the classical CGHS theory
can not be consistently quantized.
At least for the case of a compact space-time,
this problem seems to be real \cite{kawai}.
We nevertheless believe that the CGHS model still remains as a
good theoretical
laboratory in understanding the real nature of Hawking radiation.

\vspace{0.5cm}
\noindent
We thank Kazuo Fujikawa and Hikaru Kawai for useful discussions.

\newpage
\begin{center}{\bf\large Appendix}
\end{center}
\vspace{1cm}
\noindent
Before we present the algebraic
proof for the closure of the algebra (58)
under commutator, we would like to justify the chiral split
of the $Q^2$-anomaly, which is expressed by eqs. (48) and (49).

The split is by no means trivial because $\partial_{+}\psi$
($\partial_{-}\psi$)
in $Q^{(+)}_{\rm LG}$ ($Q^{(-)}_{\rm LG}$) contains
left and right movers, as one can easily see from the equation
of motion (5). One can also see that
$\partial_{+}\psi$  and $\partial_{-}\psi$ become chiral
if $\mu^2=0$. Therefore, the chiral split
(48) is absolutely correct if $\mu^2=0$, and one
may write the violation of the chiral split as
$$[\,Q_{\rm CG}^{(+)}~,~Q_{\rm CG}^{(-)}\,]~=~
i\hbar^2 \mu^2\, \Omega^{(+-2)}_{\rm CG} +O(\hbar^3)~,
\eqno{({\rm A}.1)}$$
with the most general form for $ \Omega^{(+-2)}_{\rm CG}$
(consistent with
the assumptions specified in section 4)
 $$ \Omega^{(+-2)}_{\rm CG} ~=~
\int d\sigma\,[\,\al_{1}(\,
{\cal C}^{+}\,\partial_{+}{\cal C}^{+}
+{\cal C}^{-}\,\partial_{-}{\cal C}^{-}\,)
+\al_{2}(\,{\cal C}^{+}\,\partial_{-}{\cal C}^{-}
+{\cal C}^{-}\,\partial_{+}{\cal C}^{+}\,)$$ $$
+\al_{3}\,{\cal C}^{+}{\cal C}^{-}\,(\,
Y_{+}-Y_{-}\,)\,]~. \eqno{({\rm A}.2)}$$
The term in the first parenthesis is BRST trivial as
one can easily find from the BRST transformations (47),
while the term in second parenthesis is a spatial, total
derivative which follows from the equations of motion
for ${\cal C}^{\pm}$. The consistency condition on
$\Omega^{(+-2)}_{\rm CG}$ then requires that $\al_{3}=0$.
This justifies the assumption on the chiral split (48).

Now we come to the original task. We would like to finish
the proof along the line of section 4. To this end, we
have to BRST symmetrize the residual symmetries.
Introducing a new fermionic ghost, ${\cal C}^{(+)}_{\rm f}$,
and as well as a new bosonic ghost,
${\cal C}^{(+)}_{\rm b}$, we first construct a BRST charge
for the residual symmetries:
$$
Q^{(+)}_{\rm R}~=~\int d\sigma^+ \{\,{\cal C}^{(+)}_{\rm f}\,
\partial_{+}(\rho-\phi)+{\cal C}^{(+)}_{\rm b}\,
\partial_{+}[(\,\frac{1}{2}\partial_{+}+
\partial_{+}(\rho-\phi))\,{\cal C}^{+}\,]\,\}~,\eqno{({\rm A}.3)}$$
and similarly for $Q^{(-)}_{\rm R}$, but we discuss
only the left-moving sector below (because the right-moving sector
can be treated in the exactly same way).
One, however, easily observes that the above BRST charge
does not commute with $Q^{(+)}_{\rm CG}$ (33) under Poisson
bracket. So we modify $Q^{(+)}_{\rm CG}$ as
$$
\tilde{Q}^{(+)}_{\rm CG} ~=~ Q^{(+)}_{\rm CG}+
\int d\sigma^+ \,{\cal C}^{(+)}_{\rm f}\,\ol{\cal P}^{(+)}_{b}~,
\eqno{({\rm A}.4)}$$
where  $\ol{\cal P}^{+}_{b(f)}$ is the conjugate momentum
of ${\cal C}^{+}_{\rm b(f)}$.
The algebra (58) is thus transferred to
$$\{ \tilde{Q}^{(+)}_{\rm CG}~,~Q^{(+)}_{\rm R} \}
_{\rm PB}~=~
\{ Q^{(+)}_{\rm R}~,~Q^{(+)}_{\rm R} \}_{\rm PB}~=~0~.
\eqno{({\rm A}.5)}$$
The new term
in the BRST charge $({\rm A}.4)$ can not influence the
quantum-nilpotency
property, i.e. $[\,\tilde{Q}^{(+)}_{\rm CG}~,~
\tilde{Q}^{(+)}_{\rm CG}\,]~=~0$.
Because of this,
there are two possible Schwinger terms:
$$[\,\tilde{Q}^{(+)}_{\rm CG}~,~Q^{(+)}_{\rm R}\,]~=~
i\hbar^2\, \Omega^{(+2)}_{\rm CGR}+O(\hbar^3)~, $$ $$
[\,Q^{(+)}_{\rm R}~,
{}~Q^{(+)}_{\rm R}\,]~=~
i\hbar^2\, \Omega^{(+2)}_{\rm R}+O(\hbar^3)~. \eqno{({\rm A}.6)}$$
First we consider $\Omega^{(+2)}_{\rm CGR}$ which
has to have the form
$$
\Omega^{(+2)}_{\rm CGR}~=~\int d\sigma^{+}\,\{~
{\cal C}^{(+)}_{\rm f}{\cal C}^{+}\,
\Upsilon_{\rm f}(~\partial_{+}~,~
Y_{+}~,~G_{+}~,~F^{i}_{+} ~)$$ $$
+{\cal C}^{(+)}_{\rm b}
{\cal C}^{+}\,\partial_{+}{\cal C}^{+}\,
\Upsilon_{\rm b}(~\partial_{+}~,~
Y_{+}~,~G_{+}~,~F^{i}_{+})~)~,\eqno{({\rm A}.7)}$$
where $Y_{+}~,~G_{+}~,~F^{i}_{+}~$ are defined in (46).
The consistency condition on $ \Omega^{(+2)}_{\rm CGR}$,
which can be derived from the jacobi identity
$$
2\,[\,\tilde{Q}^{(+)}_{\rm CG}~,~[\,\tilde{Q}^{(+)}_{\rm CG}~,
{}~Q^{(+)}_{\rm R}\,]\,]
+[\,Q^{(+)}_{\rm R}~,~[\, \tilde{Q}^{(+)}_{\rm CG}~,
{}~\tilde{Q}^{(+)}_{\rm CG}\,]\,]~=~0~,$$
requires at $O(\hbar^3)$ that
$$ \delta_{\rm CG}\,\Omega^{(+2)}_{\rm CGR}~=~
-\{ \tilde{Q}_{\rm CG}^{(+)}~,
{}~\Omega^{(+2)}_{\rm CGR}\}_{\rm PB}~=~0~.  \eqno{({\rm A}.8)}$$
 After some algebraic calculations, we find that the solution to
(A.8) is given by
$$\Omega^{(+2)}_{\rm CGR}~=~\delta_{\rm CG}\,\int d\sigma^+~\{~
\,{\cal C}_{\rm b}^{(+)}~[\,
\bt_{1}\,(\partial_{+})^2 {\cal C}^{+}
+\bt_{2}\,\partial_{+}Y_{+}\,{\cal C}^{+}
+\bt_{3}\,Y_{+}^2\,{\cal C}^{+}$$ $$
+\bt_{4}\, G_{+}\,{\cal C}^{+}
+\bt_{5}\,
\sum_{i=1}^{n}\,(F^{i}_{+})^2 {\cal C}^{+}\,]
+\bt_{6}\,{\cal C}_{\rm f}^{(+)}\,Y_{+}~\} ~,\eqno{({\rm A}.9)}$$ which
is BRST trivial.

As for the Schwinger term $\Omega^{(+2)}_{\rm R}$,
it has to satisfy two independent consistency conditions
$$
\delta_{\rm CG}\,\Omega^{(+2)}_{\rm R}~=~
-\{ \tilde{Q}_{\rm CG}^{(+)}~,
{}~\Omega^{(+2)}_{\rm R}\}_{\rm PB}~=~0~,$$
$$\delta_{\rm R}\,\Omega^{(+2)}_{\rm R}~=~
-\{ Q_{\rm R}^{(+)}~,
{}~\Omega^{(+2)}_{\rm R}\}_{\rm PB}~=~0~.\eqno{({\rm A}.10)}
$$
We again write the most general form for $\Omega^{(+2)}
_{\rm R}$, and go through algebraic
calculations similar to the previous ones.  One finally
finds that there is no no-trivial solution to
 (A.10).

Together with the result
on $\Omega^{(+2)}_{\rm CGR}$, we thus conclude
that the algebra (A.5) and hence (58) closes under
commutator too.

 \end{document}